\documentclass{article}

\usepackage{arxiv}

\usepackage[utf8]{inputenc} 
\usepackage[T1]{fontenc}    
\usepackage{hyperref}       
\usepackage{url}            
\usepackage{booktabs}       
\usepackage{amsfonts}       
\usepackage{nicefrac}       
\usepackage{microtype}      
\usepackage{graphicx}
\usepackage[numbers]{natbib}
\usepackage{doi}
\usepackage{enumitem}
\usepackage{makecell}
\usepackage{titlesec}
\usepackage{amsmath}
\DeclareMathOperator*{\argmax}{argmax}

\setlist{leftmargin=3.0mm}

\titlespacing\subsection{0pt}{12pt plus 4pt minus 2pt}{0pt plus 4pt minus 2pt}
\title{Coarse-grained crystal graph neural networks for reticular materials design}

\author{ \href{https://orcid.org/0000-0001-6117-5662}{\includegraphics[scale=0.07]{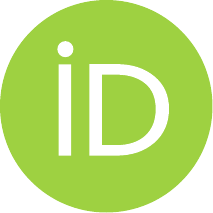}\hspace{1mm}Vadim Korolev}\thanks{\textit{Email address}: \texttt{korolev@colloid.chem.msu.ru}} \\
	Department of Chemistry\\
	Lomonosov Moscow State University\\
	Moscow 119991, Russia\\
	\And
	\href{http://orcid.org/0000-0001-8891-6862}{\includegraphics[scale=0.07]{orcid.pdf}\hspace{1mm}Artem Mitrofanov} \\
	Department of Chemistry\\
	Lomonosov Moscow State University\\
	Moscow 119991, Russia\\
}

\renewcommand{\shorttitle}{Coarse-grained crystal graph neural networks}

\begin{document}
\maketitle

\begin{abstract}
Reticular materials, including metal–organic frameworks and covalent organic frameworks, combine relative ease of synthesis and an impressive range of applications in various fields, from gas storage to biomedicine. Diverse properties arise from the variation of building units—metal centers and organic linkers—in almost infinite chemical space. Such variation substantially complicates experimental design and promotes the use of computational methods. In particular, the most successful artificial intelligence algorithms for predicting properties of reticular materials are atomic-level graph neural networks, which optionally incorporate domain knowledge. Nonetheless, the data-driven inverse design involving these models suffers from incorporation of irrelevant and redundant features such as full atomistic graph and network topology. In this study, we propose a new way of representing materials, aiming to overcome the limitations of existing methods; the message passing is performed on a coarse-grained crystal graph that comprises molecular building units. To highlight the merits of our approach, we assessed predictive performance and energy efficiency of neural networks built on different materials representations, including composition-based and crystal-structure–aware models. Coarse-grained crystal graph neural networks showed decent accuracy at low computational costs, making them a valuable alternative to omnipresent atomic-level algorithms. Moreover, the presented models can be successfully integrated into an inverse materials design pipeline as estimators of the objective function. Overall, the coarse-grained crystal graph framework is aimed at challenging the prevailing atom-centric perspective on reticular materials design.
\end{abstract}


\section{Introduction}
\label{sec:introduction}
Artificial intelligence (AI) is becoming the next game changer in materials science\cite{butler2018machine,schmidt2019recent,jablonka2020big}. Nowadays, supervised learning algorithms represent a cutting-edge tool for resolving complex structure–property relationships that determine a material’s functionality. AI methods have been utilized to examine a plethora of physicochemical parameters, including thermodynamic\cite{meredig2014combinatorial,schmidt2017predicting,bartel2020critical}, electronic\cite{lee2016prediction,zhuo2018predicting,lu2018accelerated}, mechanical\cite{evans2017predicting,mansouri2018machine,moghadam2019structure}, adsorption\cite{fernandez2014rapid,simon2015best,moosavi2020understanding}, and catalytic\cite{li2017high,zahrt2019prediction,chanussot2021open} properties. Deep learning models\cite{lecun2015deep} show state-of-the-art predictive performance by extracting crucial features from input data, e.g., chemical composition and crystal structure. Most representation schemes related to deep learning models in materials informatics require researchers to view materials from an atomistic standpoint. In particular, crystal graphs (where nodes and edges correspond to atoms and chemical bonds, respectively) enhance the AI toolset of materials scientists by means of a set of diverse graph neural networks\cite{xie2018crystal,chen2019graph,korolev2019graph,park2020developing,louis2020graph,karamad2020orbital,cheng2021geometric,choudhary2021atomistic,omee2022scalable,yan2022periodic}. Structure-agnostic neural networks\cite{jha2018elemnet,goodall2020predicting,wang2021compositionally} handle chemical composition through assigning vector representations (so-called embeddings) to the chemical elements present in a material.

Atomic-level models are highly effective in representing materials if atoms are considered the most natural choice for the elementary structural unit. In the inorganic domain, chemical subspaces can be formed through element substitution in a prototype crystal structure; the resulting material sets have been processed with predictive models in a high-throughput manner\cite{faber2016machine,ye2018deep,balachandran2018predictions}. In contrast, materials representation at a scale of molecular building units is preferable for crystalline extended structures governed by reticular-chemistry principles\cite{yaghi2003reticular,lyu2020digital}. The most prominent examples of such materials are metal–organic frameworks\cite{li1999design} (MOFs) and covalent organic frameworks\cite{cote2005porous} (COFs). In particular, MOFs are formed by linking organic molecules and metal-containing entities through coordination bonds, whereas COFs are made by stitching organic molecules through covalent bonds. Some of the synthesized compounds possess exceptional adsorption\cite{wang2018sensing,lin2019exploration} and catalytic\cite{zhu2017metal,huang2017multifunctional} properties, making them promising candidates for energy-related applications. Moreover, the modular structure of reticular materials provides a great opportunity for further tuning of relevant properties. There were recent efforts to develop specialized featurization schemes for reticular design\cite{lee2021computational,yao2021inverse,chen2022interpretable,kang2023multi,cao2023moformer}. Global descriptors (e.g., topology, volumetric attributes, and energy grids) incorporated into neural network architecture improve predictive performance and leave room for interpretability analysis. On the other hand, the aforementioned attributes constrain the scenarios where the corresponding models can be applied. The most intriguing data-driven strategy for developing functional materials—inverse design (i.e., “from property to structure”)—requires synchrony between discriminative and generative models. In particular, the input data modalities (i.e., materials representation \textit{in toto}) used in the former models should be validly reproduced by the latter ones. In this context, the incorporation of energy-grid embeddings into predictive models (e.g., MOFTransformer\cite{kang2023multi}) poses a challenge for inverse materials design owing to the limited reconstruction ability of existing algorithms. Out of one million structures, only a few zeolite shapes created by a generative adversarial network (ZeoGAN\cite{kim2020inverse}) have successfully passed all cleanup operations. Other predictive models\cite{cao2023moformer} for reticular materials directly take into consideration a framework topology, which is unknown \textit{a priori} for specified building blocks, e.g., organic linkers and metal-containing units in MOFs. Consequently, the unpredictable synthetic accessibility of frameworks (in the form of the linkers–nodes–topology triad) created by generative models has become a cornerstone for practical data-driven design of reticular materials. Recent findings confirmed concerns that most of relevant studies had overlooked: only 136 frameworks out of 1000 hypothetical structures with the highest hydrogen working capacity were identified as highly synthesizable\cite{park2022computational}.

By viewing experimental synthesis and characterization of \textit{in silico}–generated frameworks as the ultimate goal of AI-assisted materials discovery, we can formulate the main challenge in the field as follows: there exists a fundamental disparity in how synthetic chemists and AI practitioners perceive reticular materials. As a result, atomic-level and topology-aware models leverage materials modalities that are not relevant or accessible in reticular design. To address this issue, we introduce a coarse-grained crystal graph framework; the performance of neural networks that utilize sets of molecular building units as input is analyzed in terms of accuracy, energy efficiency, and transferability. In our comparative analysis, we also include the widely used architectures that are the dominant paradigm in materials representation learning; composition-based and crystal-structure–aware models are examined. In addition to material-screening applications, the feasibility of incorporating coarse-grained crystal graph neural networks into an inverse reticular design pipeline is evaluated as well.

\section{Results}
\label{sec:results}
\subsection{The coarse-grained crystal graph framework}
The modern landscape of predictive models in materials informatics\cite{reiser2022graph} is mostly shaped by neural networks built on crystal graph $(\mathcal{V},\mathcal{E})$, which is a set of vertices (atoms) $v \in \mathcal{V}$ and edges (bonds) $(u \rightarrow v) \in \mathcal{E}$. Despite impressive diversity of neural network architectures, most of them have a common foundation in the message-passing paradigm\cite{gilmer2017neural}. Vector representation ${h}_{v}$ of node $v$ is generated by propagating messages ${m}_{u \rightarrow v}$ from source nodes $u$ to destination node $v$; messages from all nodes forming neighborhood $\mathcal{N}(v)$ of node $v$ are taken into account. The message-passing procedure may be performed multiple times, and therefore node representation ${h}_{v}^{t+1}$ at step $t+1$ is computed as follows:
\begin{equation}
  {m}_{u \rightarrow v}^{t+1}=\phi({h}_{u}^{t},{h}_{v}^{t},{e}_{u \rightarrow v}^{t})
\end{equation}
\begin{equation}
  {m}_{v}^{t+1}=\rho(\left\{ {m}_{u \rightarrow v}^{t+1}, \forall u \in \mathcal{N}(v) \right\})
\end{equation}
\begin{equation}
  {h}_{v}^{t+1}=\psi({h}_{v}^{t}, {m}_{v}^{t+1})
\end{equation}
where $\phi$, $\rho$, and $\psi$ are (learnable) message, reduce, and update functions, respectively, and ${e}_{u \rightarrow v}^{t}$ is a vector representation of edge $u \rightarrow v$ at step $t$; the calculation of node features can optionally include edge features.

Our approach involves applying the message-passing method to learn representations of materials that consist of molecular building units rather than distinct atoms. The proposed framework heavily relies on subgraph neural networks\cite{alsentzer2020subgraph,sun2021sugar} and includes steps inherent in their construction. First, the strategy for sampling/selecting subgraphs needs to be explicitly stated. Following chemical intuition, we decompose MOF structures (Figure 1a) into inorganic (“metallic”) and organic components (Figure 1b); a similar approach known as the “standard simplification” algorithm\cite{barthel2018distinguishing} has been previously employed to differentiate MOF structures. Trivial subgraphs of metal atoms $(k \in \mathcal{K})$ (Figure S1) are composed of single nodes in the crystal graph; intermetallic bonds are ignored. Another set of subgraphs $(l \in \mathcal{L})$ is formed by the connected components of nonmetal atoms. Domain heuristics enable the development of alternative partition schemes that deconstruct MOFs into organic linkers and inorganic substructures, commonly known as secondary building units. Our preliminary experiments showed that those methods are not universally applicable and cannot extract molecular building units for many synthesized MOFs. For instance, 94.5\% and 71.8\% of entities from the Computation-Ready, Experimental Metal-Organic Framework (CoRE MOF) database\cite{chung2014computation,chung2019advances} are correctly processed by the MOFid\cite{bucior2019identification} and moffragmentor\cite{jablonka2023ecosystem} packages, respectively. On the other hand, our implemented scheme achieved identification of metal centers and organic linkers in 99.0\% of structures. Decomposition rates of other databases are similar (Figure S2). In the case of COFs, we apply a similar methodology, with the only variation being the composition of inorganic nodes in terms of chemical elements; here boron and silicon are also taken into account (Figure S1). Consequently, the current stage of development does not support adequate processing of a diverse family of metal-free COF structures within the proposed framework.

The next step is to define subgraph representations. Crystal graph neural networks appear to be the most relevant reference in this context; during training, learnable atomic embeddings are commonly generated in input layers of a model. In contrast, we rely on predefined vectors to characterize both the inorganic $\mathcal{K}$ subset and organic $\mathcal{L}$ subset. Specifically, the mol2vec model\cite{jaeger2018mol2vec} is implemented to represent molecular fragments, and matscholar embeddings\cite{weston2019named} serve as features for single-node metal subgraphs (Figure 1c).

The considered sets of entities ($\mathcal{K}$ and $\mathcal{L}$) form bipartite graph $(\mathcal{K},\mathcal{L},\mathcal{E})$ by design; each edge in the graph connects species of different types, i.e., an organic linker to a metal center and \textit{vice versa}. Unfortunately, reticular design does not provide prior knowledge of the specific bonds that will form in an experimental crystal structure. To eliminate arbitrariness from the selection of initial connectivity, we postulate every potential edge thereby forming complete bipartite graph $(\mathcal{K},\mathcal{L},\tilde{\mathcal{E}})$ called the coarse-grained crystal graph. The term refers to the coarse-graining modeling approach, which involves simplification of complex atomic systems, e.g., reducing the number of degrees of freedom by representing groups of atoms as a single pseudoatom. Similar techniques have been applied to quantify MOF diversity\cite{nicholas2020understanding,nicholas2021visualization} (in an unsupervised manner) and to establish structure–property relationships in a specific class of hybrid materials\cite{beaulieu2023coarse} (zeolitic imidazolate frameworks). To the best of our knowledge, this study is the first to use the message-passing paradigm in order to model reticular materials under the coarse-grained regime.

Coarse-grained crystal graphs can be easily integrated into graph neural networks designed for heterogeneous graphs. To handle structure–property relationships in reticular materials, we implement a model with simple architecture (Figure 1d), hereafter designated as the coarse-grained crystal graph neural network (CG\textsuperscript{2}-NN). Predefined embeddings initiate the sequential update of node representations through the message-passing procedure occurring in three interaction blocks; each block includes a convolutional layer, layer normalization\cite{ba2016layer}, and nonlinearity: the Exponential Linear Unit\cite{clevert2015fast} (ELU). Owing to the universal nature of the proposed structure representation, a wide range of convolutional operations can be integrated into the interaction block. In this study, we conducted experiments using four message-passing techniques: Graph Convolutional Network layer\cite{kipf2016semi} (CG\textsuperscript{2}-GCN), SAmple and aggreGatE layer\cite{velivckovic2017graph} (CG\textsuperscript{2}-SAGE), Graph ATtention layer\cite{hamilton2017inductive} (CG\textsuperscript{2}-GAT), and Graph ATtention layer with a universal approximator attention function\cite{brody2021attentive} (CG\textsuperscript{2}-GATv2). It should also be emphasized that the aggregation phase of message passing is obviously influenced by the specific topology of the coarse-grained crystal graph (Equation 2). Because each linker has a neighborhood comprising all metal centers and each metal center has a neighborhood composed of all linkers, this phase looks as follows:
\begin{equation}
  {m}_{k}^{t+1}=\rho(\left\{ {m}_{l \rightarrow k}^{t+1}, \forall l \in \mathcal{L} \right\})
\end{equation}

\begin{equation}
  {m}_{l}^{t+1}=\rho(\left\{ {m}_{k \rightarrow l}^{t+1}, \forall k \in \mathcal{K} \right\})
\end{equation}

\begin{figure}[ht]
  \centering
  \includegraphics[width=16.8cm]{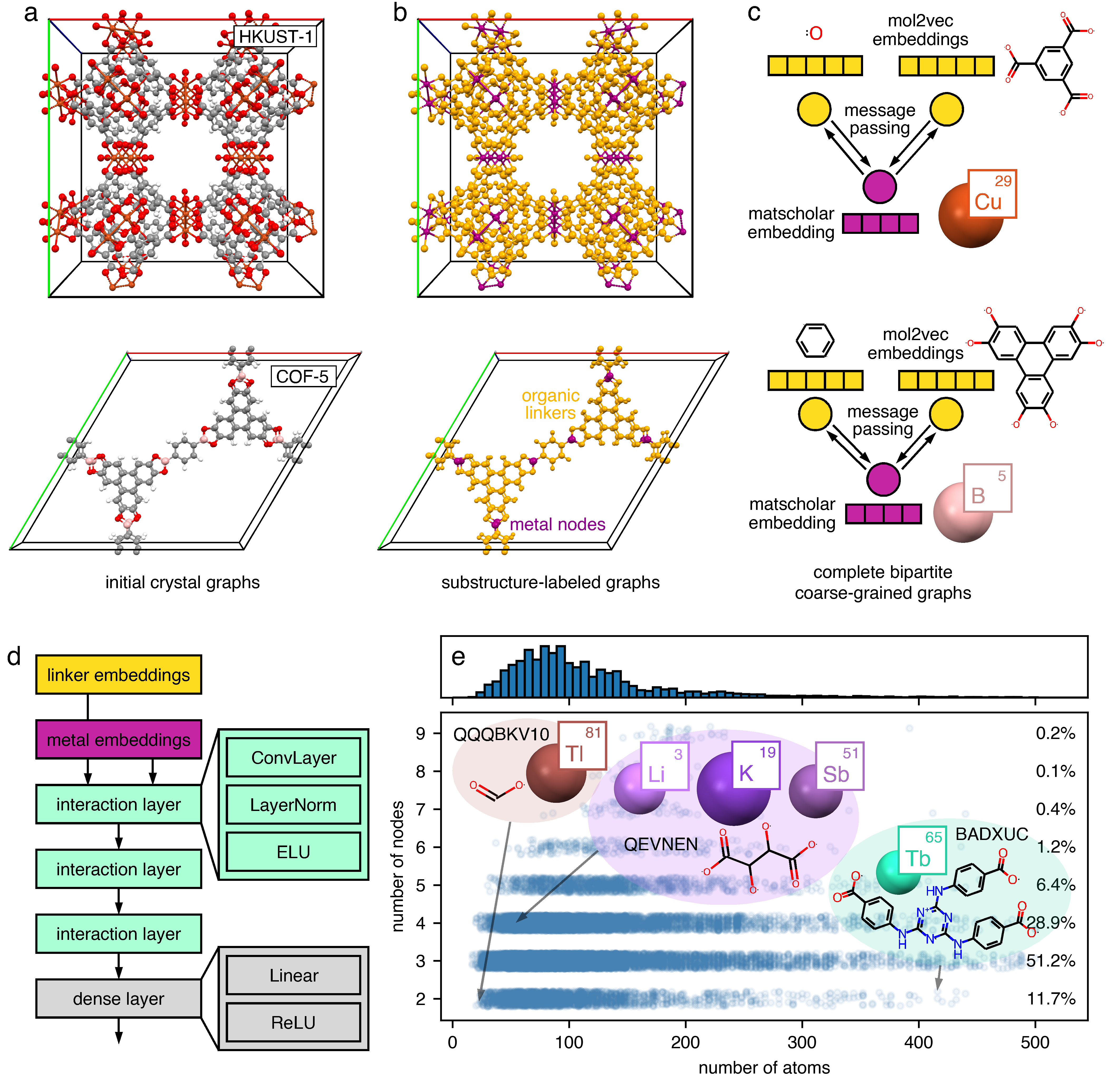}
  \caption{The coarse-grained crystal graph framework. The overview of data processing and scalability. (a) Original crystal structures of notable reticular materials: metal–organic framework HKUST-1 and covalent organic framework COF-5. Atoms are colored based on their chemical element. (b) The crystal graphs of HKUST-1 and COF-5 are colored based on their substructure type. The metal and nonmetal atoms are violet and yellow, respectively. (c) The scheme of the construction of a coarse-grained crystal graph. (d) A schematic diagram of the coarse-grained crystal graph neural network architecture implemented in this study. (e) The number of nodes in the coarse-grained crystal graph as a function of the number of atoms in the corresponding unit cell. The top panel contains the corresponding distribution of the number of atoms.}
  \label{fig:fig1}
\end{figure}

The calculation of graph level representation is based on the following readout function:
\begin{equation}
  {h}_{g} = \frac{1}{|\mathcal{K}|}\sum_{k \in \mathcal{K}} {h}_{k} + \frac{1}{|\mathcal{L}|}\sum_{l \in \mathcal{L}} {h}_{l}
\end{equation}
Finally, the output of the CG\textsuperscript{2}-NN is obtained by passing graph level embedding ${h}_{g}$ through a dense layer.

In terms of message-passing scalability, the coarse-grained crystal graph framework is expected to outperform the full crystal graph. The Quantum MOF (QMOF) database\cite{rosen2021machine,rosen2022high} contains structures with hundreds of atoms per primitive cell (Figure 1e). It should be noted that the initial set of candidate materials was compiled considering the limitations of high-throughput density functional theory (DFT) calculations. The structures in other experimental MOF subsets, such as the Crystal Structure Database (CSD) MOF Collection\cite{li2021launch} and CoRE MOF database, can contain as many as 10,000 atoms (Figure S3). In contrast, the coarse-grained crystal graph has a maximum of 9, 17, or 35 nodes in the three databases mentioned above; five or fewer vertices are common in most coarse-grained crystal graphs (98.1\%, 91.1\%, and 92.8\% of cases). Probably the most illustrative quantity—the average ratio of nodes in a crystal graph to nodes in the coarse-grained crystal graph—equals 35.3, 125.6, and 101.8 in the QMOF database, CSD MOF Collection, and CoRE MOF database, respectively. Nonetheless, the impressive scalability is only partially attributable to coarse-graining molecular subgraphs. Another factor is the occurrence of multiple identical subgraphs in the primitive cell; the coarse-grained crystal graph is intentionally free of duplicates, which means that a specific ratio of building units in the original reticular structure is disregarded. As a result, CG\textsuperscript{2}-NNs are likely to show worse predictive performance as compared to crystal graph neural networks. The following sections primarily address the efficiency-vs-accuracy dilemma of the proposed computational framework.

\subsection{Predictive performance of CG\textsuperscript{2}-NNs}
To evaluate the predictive performance of CG\textsuperscript{2}-NNs, we consider a diverse set of practically important MOF properties, including the band gap, thermal decomposition temperature, heat capacity, and Henry coefficients of eight gases (N\textsubscript{2}, O\textsubscript{2}, Kr, Xe, CH\textsubscript{4}, CO\textsubscript{2}, H\textsubscript{2}O, and H\textsubscript{2}S); the tasks in question are all in a regression setting. In addition to the proposed reticular-specific architecture, several other neural networks that are composition-based and (crystal-)structure-aware are benchmarked as well. The algorithm designated as Representation Learning from Stoichiometry\cite{goodall2020predicting} (Roost) and a model within the framework of Wyckoff Representation regression\cite{goodall2022rapid} (Wren) form the first group. It is interesting to note that Wren incorporates Wyckoff representations in addition to stoichiometry, thus introducing the concept of coordinate-free coarse graining into materials discovery. We categorize this approach as composition-based because the model does not directly incorporate the full crystal graph. Another group of methods includes four crystal graph neural networks: Crystal Graph Convolutional Neural Network\cite{xie2018crystal} (CGCNN), MatErials Graph Network\cite{chen2019graph} (MEGNet), Global ATtention-based Graph Neural Network with differentiable group normalization and residual connection\cite{omee2022scalable} (DeeperGATGNN), and Atomistic Line Graph Neural Network\cite{choudhary2021atomistic} (ALIGNN). Aside from ”general-purpose” structure-aware architectures, a few recent articles introduced neural networks that integrate domain knowledge related to reticular chemistry; MOFNet\cite{chen2022interpretable}, MOFTransformer\cite{kang2023multi}, and MOFormer\cite{cao2023moformer} are worth mentioning. MOF-related models are not a part of our benchmark analysis; the original studies provide an overall picture of accuracy by comparing with crystal graph neural networks, e.g., CGCNN.

\begin{figure}[t!]
  \centering
  \includegraphics[width=16.0cm]{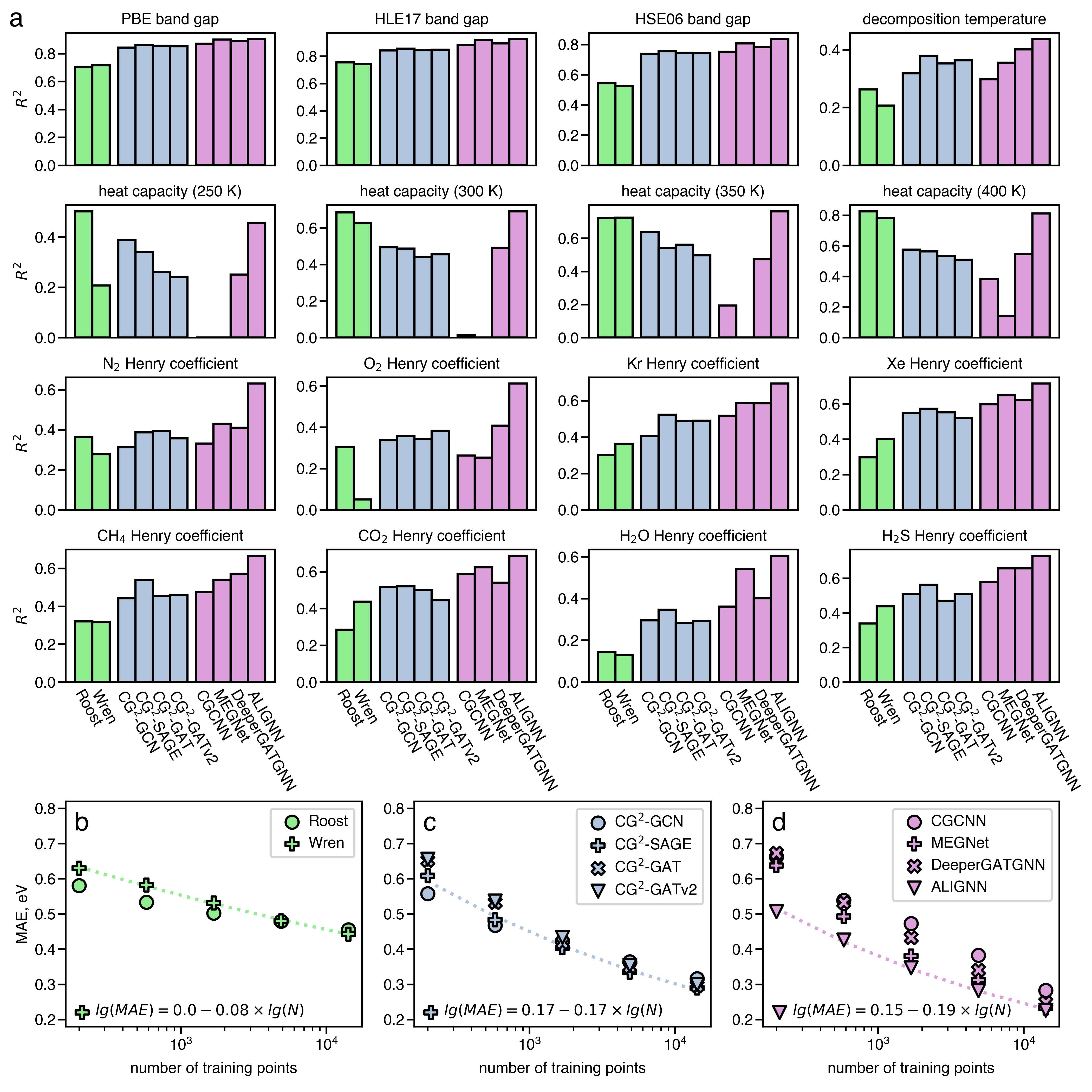}
  \caption{Predictive performance of three model classes. (a) Coefficients of determination (${R}^{2}$) of composition-based, coarse-grained crystal graph, and crystal-structure–aware neural networks are highlighted in green, blue, and violet, respectively. (b–d) Mean absolute error (MAE) in the band gap prediction as a function of training dataset size.}
  \label{fig:fig2}
\end{figure}

Despite acknowledging the criticisms of using the coefficient of determination as a primary predictive performance measure\cite{alexander2015beware}, we still find it useful for identifying general trends (Figure 2a). To ensure completeness, we provide the corresponding mean absolute error (MAE) and root mean square error (RMSE) values for all tasks (Figures S4 and S5). From the metrics, we can conclude that CG\textsuperscript{2}-NNs are surprisingly effective in predicting a DFT band gap. A coefficient of determination for the best models (CG\textsuperscript{2}-SAGE) proved to be 0.86, 0.85, and 0.76 at three levels of theory (see details in the Methods section). In comparison, the best crystal graph neural networks (ALIGNN) have a coefficient of determination of 0.90, 0.92, and 0.84 for the same tasks. The accuracy of composition-based models was found to be substantially lower, with a coefficient of determination of 0.72, 0.75, and 0.54, respectively. The prediction of thermal decomposition temperature presents a challenge for all implemented models. In particular, CG\textsuperscript{2}-SAGE and ALIGNN reached a coefficient of determination of 0.38 and 0.44, respectively. We can speculate that the low accuracy is mainly due to relatively high uncertainty of determining a target value from thermogravimetric analysis (TGA) data. A typical rounding step of 25 °C (applied to decomposition temperature values\cite{healy2020thermal}) is comparable to MAE of predictive models: 50 and 47 °C in the case of CG\textsuperscript{2}-SAGE and ALIGNN, respectively. Therefore, the semiquantitative agreement between experimental and predicted values can hardly be enhanced without expanding the set of target values and improving the resolution limit of TGA data. Nonetheless, CG\textsuperscript{2}-NNs and structure-aware models were found to have similar coefficients of determination, 0.32–0.38 vs. 0.30–0.44. The next endpoint, thermal capacity, poses a different challenge for predictive models. Unexpected model rankings and ineffective training were caused by a shortage of data points (only 214) in the dataset. Composition-based Roost and structure-aware ALIGNN showed comparable performance, with an average coefficient of determination of 0.68 for four temperature values. On the other hand, CGCNN and MEGNet, both incorporating crystal graph data, demonstrated nearly zero predictive performance. CG\textsuperscript{2}-NNs manifested moderate accuracy; CG\textsuperscript{2}-GCN achieved the highest average coefficient of determination of 0.52 among these models. Finally, Henry coefficients are examined in the analysis (Figure 2a); the following metrics were calculated by averaging eight endpoint-wise quantities. ALIGNN with a coefficient of determination of 0.67 significantly outperformed all other models; the second best MEGNet showed a coefficient of determination of 0.54. CG\textsuperscript{2}-SAGE achieved a coefficient of determination of 0.48, surpassing structure-aware CGCNN (0.46). Both composition-based models, Roost and Wren, showed a limited predictive ability with an average coefficient of determination of 0.29 and 0.30, respectively. The difference in predictive performance between ALIGNN and the other models may be due to the significance of the specific geometry of adsorption sites in MOF; the explicit inclusion of three-body interaction terms in ALIGNN enables precise handling of atomic environments. To sum up, CG\textsuperscript{2}-NNs manifested accuracy comparable to simplistic structure-aware models, e.g., CGCNN, and generally outperformed composition-based neural networks.

\begin{figure}[t!]
  \centering
  \includegraphics[width=16.7cm]{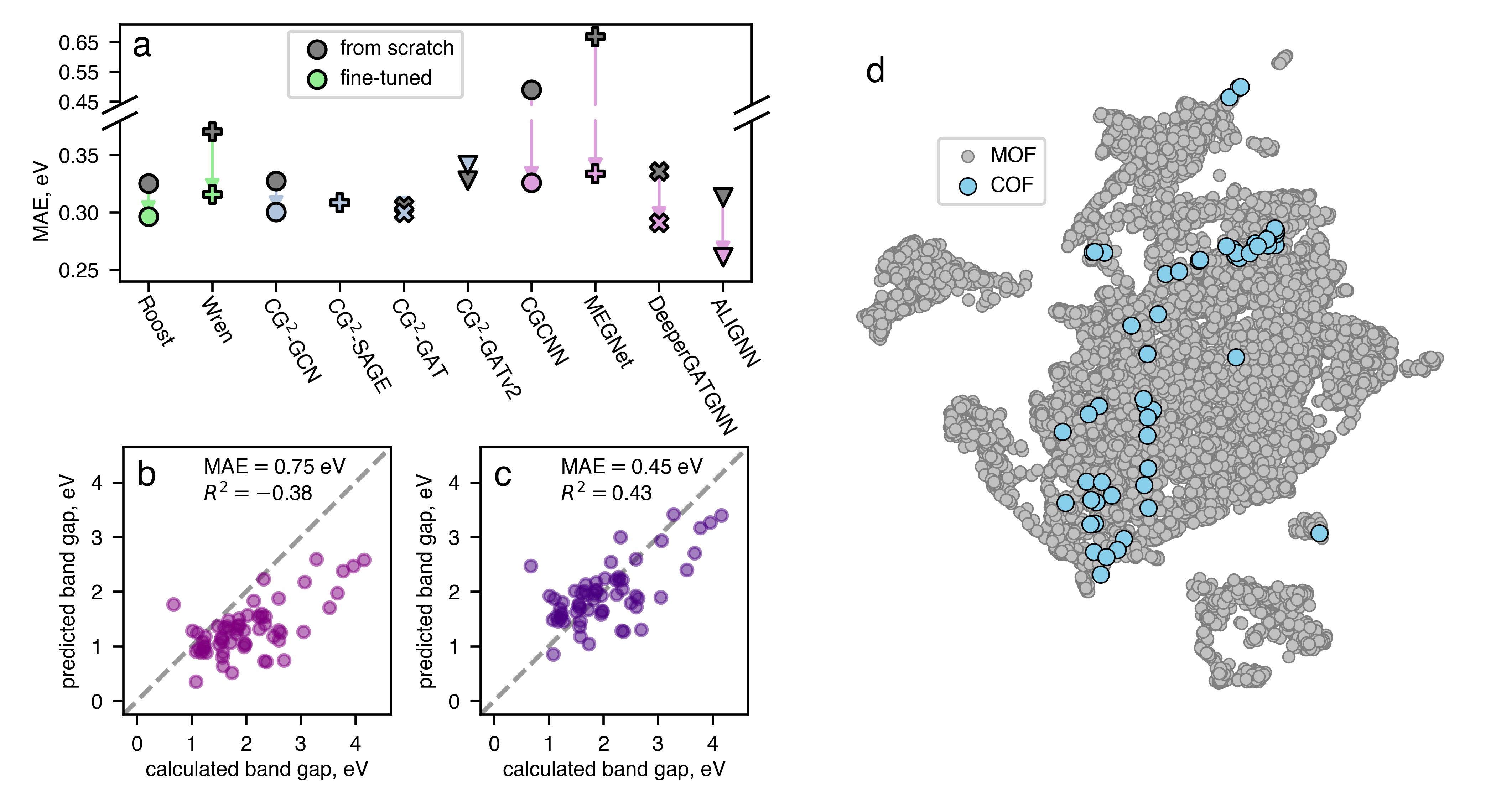}
  \caption{Cross-domain (from metal–organic frameworks [MOFs] to covalent organic frameworks [COFs]) transferability of coarse-grained crystal graph neural networks. (a) Mean absolute error (MAE) in the band gap prediction. MAE values for models pretrained on MOF data are shown in color, while the models exclusively trained on COF data are gray. (b) A scatter plot of calculated and predicted values of the band gap. (c) A scatter plot of calculated and predicted values of the band gap; linear scaling to minimize MAE is applied. (d) The two-dimensional projection of linker chemical space is produced within the Uniform Manifold Approximation and Projection (UMAP) algorithm from mol2vec embeddings; the linkers from the Quantum MOF database and from the subset of CURATED COFs are gray and blue, respectively.}
  \label{fig:fig3}
\end{figure}

Next, we analyze the scalability of the proposed computational framework with respect to the size of the training dataset (Figure 2b–d) while focusing on the PBE band gap; the models from each of the above-mentioned classes (composition-based, coarse-grained, and structure-aware) are examined. We approximate how MAE depends on training-dataset size $N$ using a linear fit on a logarithmic scale for three models: Wren, CG\textsuperscript{2}-SAGE, and ALIGNN. As follows from fitted coefficients $a$ and $b$ in equation $\lg MAE = a + b \lg N$ (Figure 2b–d), CG\textsuperscript{2}-SAGE and ALIGNN showed similar behavior, whereas the increase in training-dataset size led to a much smaller boost in performance for Wren. At the same time, most structure-aware models were outperformed by Wren and another composition-based model (Roost) under the small-data regime (200 points in the training dataset); the observed tendency for heat capacity prediction was reproduced here.

The lack of data for the endpoint of interest can be addressed by leveraging advanced techniques, including transfer learning\cite{yamada2019predicting,jha2019enhancing,gupta2021cross} and self-supervised learning\cite{suzuki2022self,korolev2023accurate}. To assess potential usefulness of applying CG\textsuperscript{2}-NNs in conjunction with one of such algorithms, we consider the task of band gap prediction across different classes of reticular materials: the models pretrained on the MOF band gap are used to evaluate the same property in COFs (Figure 3a). The models trained from scratch, i.e., without pretraining, serve as a baseline. CG\textsuperscript{2}-GAT and ALIGNN showed the lowest MAE of 0.31 eV; most of the models performed at a similar level. In contrast, two structure-aware models—CGCNN and MEGNet—failed to quantitatively reproduce the target value, judging by the thermal capacity prediction of MOFs (Figure 2a). On the other hand, the transfer-learning technique resulted in the largest decrease in MAE (i.e., increase in accuracy) for these models: 0.16 and 0.34 eV, respectively. The lowest MAE of 0.26 eV was achieved by the fine-tuned ALIGNN model; DeeperGATGNN with MAE of 0.29 eV was the second-best model owing to the moderate performance improvement (0.05 eV) associated with fine-tuning. Pretraining with MOF data had a minimal effect on the accuracy of all CG\textsuperscript{2}-NNs. Minor benefits of transfer learning may stem from incorporating pretrained embeddings into the proposed neural network architecture; at the concept level, this approach is essentially equivalent to freezing the weights of the input layer. The mol2vec (matscholar) embeddings are learned here by exploring vast chemical space, which extends beyond the organic linkers (metal centers) in reticular materials. In other words, the generalizability of low-level representations of molecular building units allows for the description of both MOFs and COFs from scratch. The band gap of COFs can be semiquantitatively reproduced by CG\textsuperscript{2}-GAT trained on MOF data, without additional fine-tuning; only a linear scaling of the model outputs is needed (Figure 3b,c). From another perspective, common chemical space is formed by two-dimensional projections of organic linkers in MOFs and COFs; these projections are generated within the Uniform Manifold Approximation and Projection\cite{mcinnes2018umap} (UMAP) approach from mol2vec embeddings (Figure 3d).

\begin{figure}[t!]
  \centering
  \includegraphics[width=16.5cm]{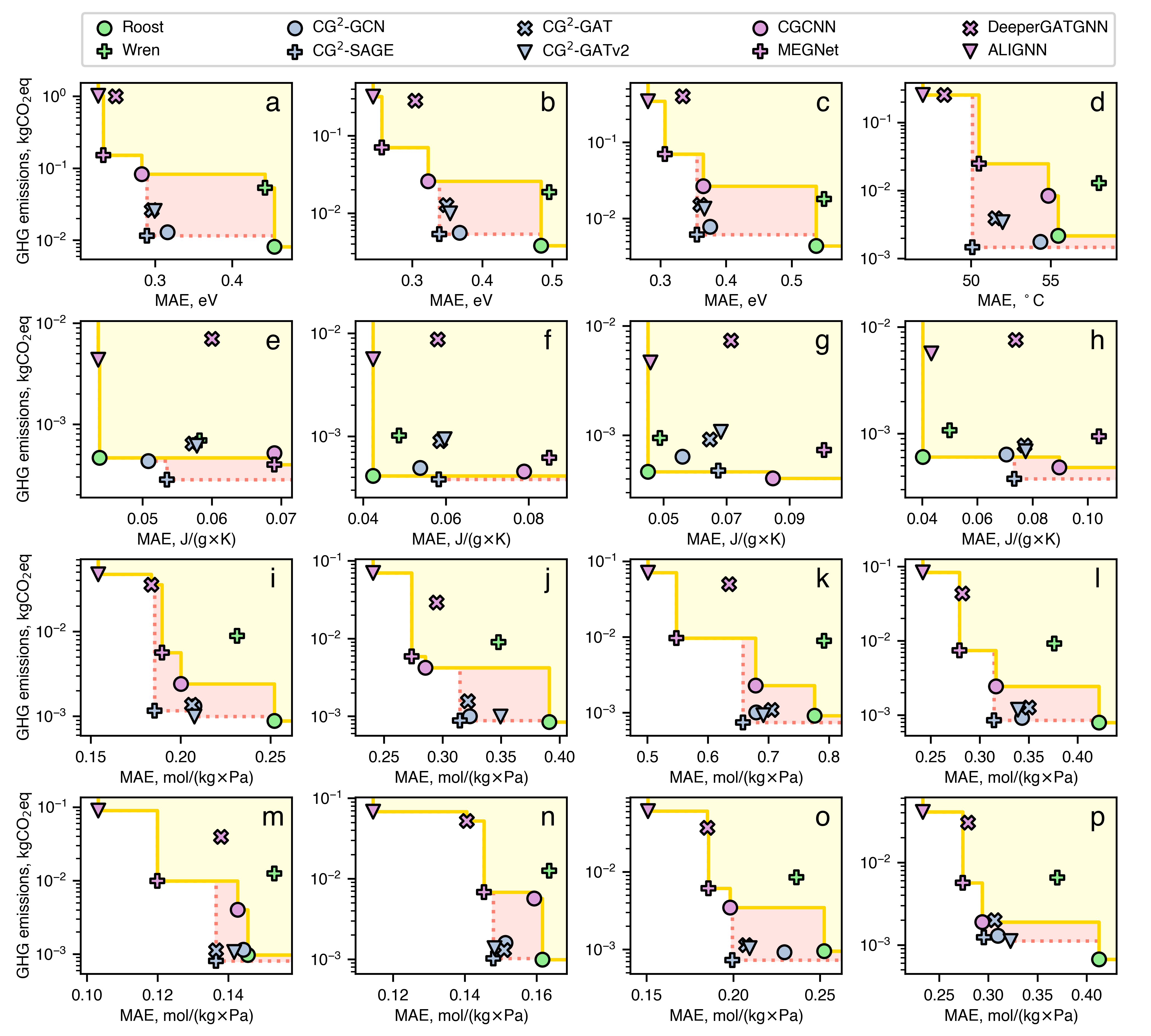}
  \caption{Interplay between predictive performance and energy efficiency of composition-based, coarse-grained crystal graph, and crystal-structure–aware models. The following endpoints are considered: (a) the PBE band gap, (b) the HLE17 band gap, (c) the HSE06 band gap, (d) thermal decomposition temperature, (e) heat capacity at 250 K, (f) heat capacity at 300 K, (g) heat capacity at 350 K, (h) heat capacity at 400 K, (i) the N\textsubscript{2} Henry coefficient, (j) the O\textsubscript{2} Henry coefficient, (k) the Kr Henry coefficient, (l) the Xe Henry coefficient, (m) the CH\textsubscript{4} Henry coefficient, (n) the CO\textsubscript{2} Henry coefficient, (o) the H\textsubscript{2}O Henry coefficient, and (p) the H\textsubscript{2}S Henry coefficient. For illustrative purposes, the areas of MAE vs. GHG emission space accessible by different models are highlighted in two colors: yellow regions correspond to the set of composition-based and structure-aware models, whereas red fields reflect how the Pareto front is reshaped by the introduction of CG\textsuperscript{2}-NNs.}
  \label{fig:fig4}
\end{figure}

\subsection{The carbon footprint of training CG\textsuperscript{2}-NNs}
The main focus of the materials informatics community has been on enhancing predictive performance of existing neural network architectures\cite{probst2023aiming}, paying little attention to other critical aspects such as computational efficiency, explainability, transferability, and scalability. As shown in our recent study\cite{korolev2023carbon}, an excessive focus on model accuracy has led to exponential growth of trainable parameters and greenhouse gas (GHG) emissions from model training. Taking into account the decent predictive ability and high scalability (in terms of input data size) of CG\textsuperscript{2}-NNs, we also estimate the carbon footprint of our models in the hope that the presented framework can tackle the accuracy–efficiency dilemma in materials property prediction. In Figure 4, the interplay between GHG emissions (measured in kilograms of carbon dioxide equivalents, kgCO\textsubscript{2}eq) and predictive performance (expressed in terms of MAE) is analyzed for MOF-related models from the previous section. It is worth noting that the carbon footprint of the model lifecycle is solely determined by the electrical energy consumption of the hardware in use. Accordingly, GHG emissions can provide a description of energy efficiency, in addition to quantifying environmental impacts. In contrast to the issue of ranking models by the accuracy objective, here we deal with a set of so-called nondominated solutions, i.e., models that provide a tradeoff between target quantities. For instance, ALIGNN ended up in the low-error section of the Pareto front in all cases, except for thermal-capacity prediction. The composition-based Roost model was the only exception, because it showed high predictive performance under the small-data regime. Roost and CG\textsuperscript{2}-SAGE were found to dominate the low-emission region of the Pareto front, thereby confirming the impressive computational efficiency of the proposed framework. As we can see, CG\textsuperscript{2}-NNs offer nearly state-of-the-art performance at a fraction of the computational cost of previous algorithms. For instance, the PBE band gap MAE of CG\textsuperscript{2}-SAGE can be reduced by 20\% (22\%) with MEGNet (ALIGNN), causing GHG emissions to increase 13-fold (89-fold); similar estimates apply to other endpoints as well. Therefore, the coarse-grained crystal graph concept offers a strong alternative to current dominant methods for representing reticular materials if predictive performance and energy efficiency are equally important.

\subsection{Inverse materials design accompanied with CG\textsuperscript{2}-NNs}
CG\textsuperscript{2}-NNs are readily accessible as discriminative algorithms for \textit{in silico} high-throughput screening of reticular materials. Nevertheless, the presented computational framework can be integrated into inverse design pipelines as well. As a proof-of-concept assessment, we investigate maximizing hydrogen storage capacity in MOFs; the optimization task is given by
\begin{equation}
  {x}^{*} = \argmax_{x \in \mathcal{X}} f(x)\
\end{equation}
where $\mathcal{X}$ is design space represented within the coarse-grained crystal graph framework, $f(x)$ is an objective function (hydrogen storage capacity) approximated by a CG\textsuperscript{2}-NN, and ${x}^{*}$ is an optimal set of metal centers $\mathcal{K}$ and organic linkers $\mathcal{L}$ encoded as complete bipartite graph $(\mathcal{K},\mathcal{L},\tilde{\mathcal{E}})$. Using IRMOF-20\cite{rowsell2006effects} as a prototype, we apply an iterative evolutionary procedure to modify the linker (thieno[3,2-b]thiophene-2,5-dicarboxylic acid) and maximize the target property; the metal center (zinc) and another linker (oxygen atom) are kept unchanged. In other words, we move from a global optimization problem to a local maximization one; our intention is to confine the design space to the region surrounding the original structure. At each optimization step (Figure 5a), SELF-referencIng Embedded Strings\cite{krenn2020self} (SELFIES) representation of a molecular building block undergoes one of three operations: single-character addition, deletion, or replacement. SELFIES strings are 100\% chemically robust by design, satisfying valence-bond rules under any mutation, but other specific requirements should also be fulfilled by a potential linker. In particular, the “linker-likeness” and synthetic-accessibility filters are implemented in the pipeline (see details in the Methods section). The linkers that meet the above criteria are passed through CG\textsuperscript{2}-SAGE as part of the coarse-grained crystal graph for predicting hydrogen storage capacity. Next, we evaluate the uncertainty in predictions using the deep-ensemble method\cite{lakshminarayanan2017simple}. The variance of model outputs calculated for a set of neural networks may serve as a reliable measure of epistemic uncertainty, which indicates limitations of a model in reproducing structure–property relationships outside its domain of applicability. As we can see in Figure 5b, narrower prediction intervals in terms of percentiles (defined by means of quantile regression\cite{korolev2022universal}) are associated with lower standard deviation of model ensemble outputs. Accordingly, in-domain MAE decreases with lowering the upper bound of standard deviation (Figure 5c); the structure is considered to be inside the domain of applicability if the uncertainty measure is below the predefined threshold. In the context of optimization problems, the ranking capability of a model appears to be a key parameter; the implied threshold value (2.5 g/l) allowed to reach a Pearson correlation coefficient of 0.86 (Figure 5c). Finally, hydrogen storage capacity of the generated linker must exceed the lowest one in the previous population.

\begin{figure}[t!]
  \centering
  \includegraphics[width=16.6cm]{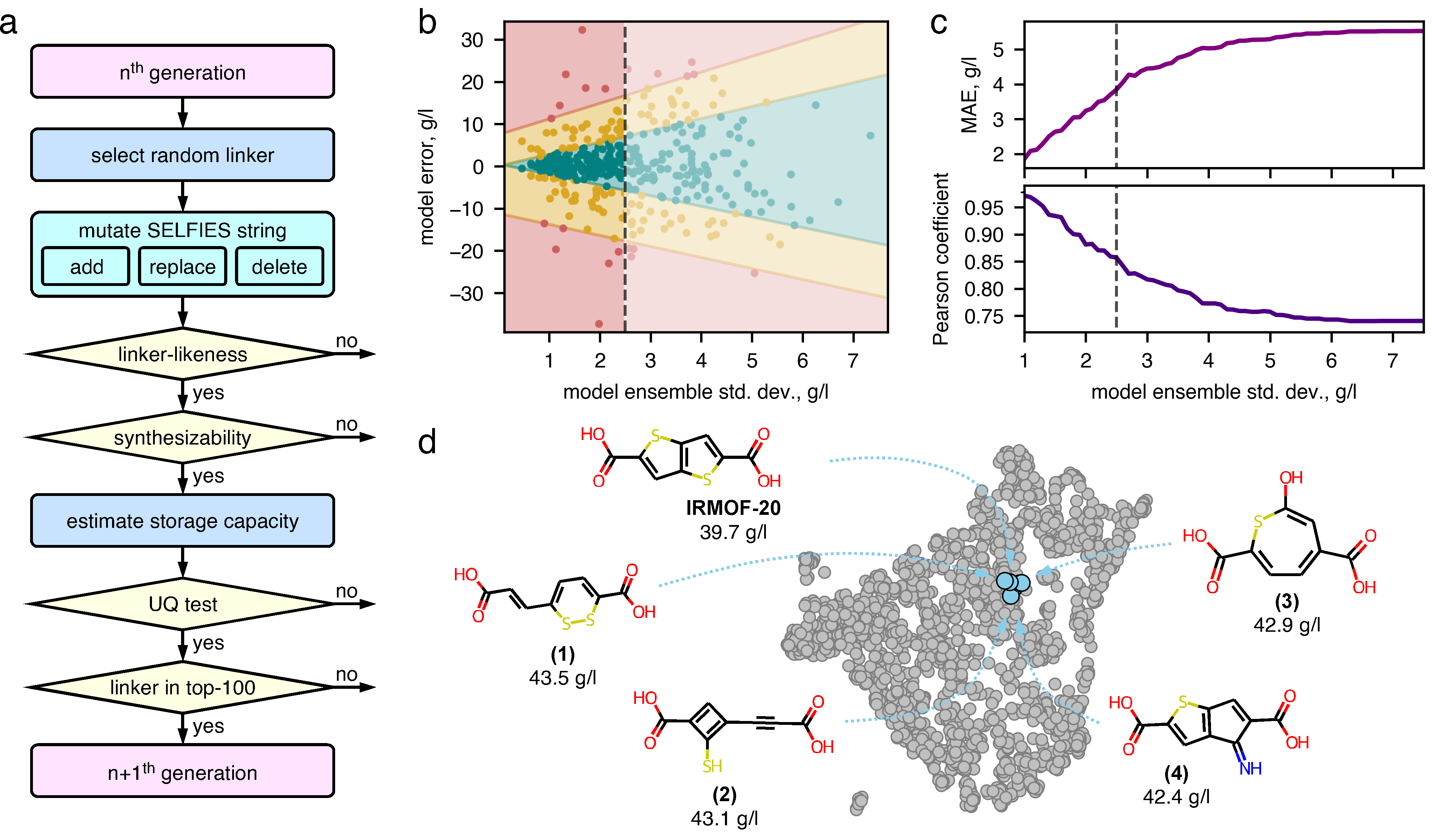}
  \caption{Inverse reticular design driven by coarse-grained crystal graph neural networks. (a) The schematic diagram of the linker optimization workflow. SELFIES is an abbreviation of SELF-referencIng Embedded Strings; UQ stands for uncertainty quantification. (b) Predictive error as a function of model ensemble standard deviation. The colored areas correspond to the predictive intervals estimated by quantile regression: within one standard deviation (green), within two standard deviations (orange), and over two standard deviations (red). (c) In-domain mean absolute error (MAE) and the Pearson correlation coefficient as a function of model ensemble standard deviation. The threshold value of standard deviation (2.5 g/l) is indicated by the vertical line. (d) The initial linker (taken from IRMOF-20) and four generated molecules with the highest hydrogen storage capacity. The two-dimensional projection of linker chemical space is produced within the Uniform Manifold Approximation and Projection (UMAP) algorithm from mol2vec embeddings; the linkers from MOFs used for training objective function predictors are shown in gray.}
  \label{fig:fig5}
\end{figure}

Additionally, the top-performing linkers undergo filtering based on their 3D structure. Rigid molecules with antiparallel binding sites, e.g., carboxylic groups, are regarded as potential precursors. The set consists of molecules that possess both the original scaffold (linker \textbf{4}) and a new one (linkers \textbf{1}, \textbf{2}, and \textbf{3}); they are all located in the vicinity of the initial compound (thieno[3,2-b]thiophene-2,5-dicarboxylic acid) in the chemical space (Figure 5d). The parental MOF structure (IRMOF-20) showed nearly record-breaking performance (experimental volumetric capacity of 51 g/l\cite{ahmed2017balancing}); therefore, the generated linkers only slightly enhance the predicted target property value (up to 3.8 g/l). Expanding the design space of the optimization problem, e.g., by loosening the uncertainty-quantification criterion, may improve the quality of found solutions; simultaneous optimization of a metal center and organic linker achieves the same goal. Nonetheless, the molecules that were identified showed potential as building units for MOFs with high hydrogen working capacity. The presented inverse-design pipeline provides outputs that can be readily utilized to predict MOF synthesis parameters\cite{luo2022mof,zheng2023chatgpt}.

\section{Discussion}
As demonstrated in the Results section, CG\textsuperscript{2}-NNs are able to compete in predictive performance with models incorporating information on atomic connectivity. This finding suggests that the underlying topology of a reticular material can be learned during model training. We intentionally examine datasets containing experimentally resolved MOFs and COFs; the relationship between a set of molecular building blocks and their self-assembled structure is usually straightforward. On the contrary, widely used databases of hypothetical structures expand the reticular materials genome by providing multiple structures of various topologies for a specific organic linker and secondary building unit. Obviously, CG\textsuperscript{2}-NNs will fail to distinguish such structures, although most of \textit{in silico}–generated MOFs are thermodynamically inaccessible and, as a result, are not of practical interest. The synthesized MOFs that exhibit polymorphism\cite{aulakh2015importance,widmer2019rich} will cause the same issue. Incorporating global state attributes, e.g., temperature and pressure, into CG\textsuperscript{2}-NN architecture can potentially overcome the challenge of polymorphic phase transitions.

By eliminating topology from explicit consideration, we address the key shortcoming of current approaches to the creation of reticular materials. The entire design space is defined by the linkers–nodes–topology triad, but the optimization task is constrained by reticular chemistry principles: an experimentally observed topology is determined by molecular building units and is not an independent variable. The issue has been mostly ignored in previous studies\cite{lee2021computational,zhou2022inverse,comlek2023rapid}, whereas the energy ranking of polymorphs requires considerable computing resources\cite{park2022computational,lim2021finely}. We anticipate that the coarse-grained crystal graph framework will shift the community’s focus from navigating inaccessible regions of the reticular materials genome to improving methods for predicting synthesis conditions.

\section{Conclusion}
The coarse-grained crystal graph framework is intended to make data-driven design of reticular materials more accessible for synthetic chemists. The efficiency of the presented approach is achieved via integration of relevant domain knowledge into the neural network architecture. Specifically, pretrained embeddings are used to represent organic linkers and metal centers, thereby demonstrating the impact of previous community efforts on AI tools for materials design. We hope that this research will not end here but rather bolster further enhancement of predictive models.

\section{Methods}
\subsection{Datasets}
The QMOF database\cite{rosen2021machine,rosen2022high} served as a source for MOF crystal structures and their corresponding band gap values obtained at three levels of theory: generalized gradient approximation (GGA), meta-GGA, and screened-exchange hybrid GGA; the density functionals used were Perdew–Burke–Ernzerhof\cite{perdew1996generalized} (PBE), High Local Exchange 2017\cite{verma2017hle17} (HLE17), and Heyd–Scuseria–Ernzerhof\cite{krukau2006influence} (HSE06), respectively. The datasets consisted of 20237 (PBE endpoint), 10664 (HLE17), and 10718 (HSE06) entities. Another collection of neural networks was trained by means of a subset of structures (3038 points) from the CoRE MOF 2019 database and their corresponding decomposition temperatures extracted from the TGA data by Nandy et al.\cite{nandy2022mofsimplify} For the heat capacity prediction, we utilized the values obtained within the harmonic approximation at the GGA level of theory\cite{moosavi2022data}; 214 MOFs from the CoRE MOF 2019 and QMOF database were examined. Henry coefficients of eight gases—N\textsubscript{2}, O\textsubscript{2}, Kr, Xe, CH\textsubscript{4}, CO\textsubscript{2}, H\textsubscript{2}O, and H\textsubscript{2}S—computed via grand canonical Monte Carlo (GCMC) simulations and the corresponding structures from the QMOF database (1431, 1552, 1297, 1205, 1268, 1538, 1482, and 1352 compounds, respectively) were taken from the dataset presented by Jablonka et al.\cite{jablonka2023ecosystem} PBE band gap values\cite{mourino2023search} of 61 compounds (from the CURATED COFs database\cite{ongari2019building}) containing boron or silicon atoms were utilized to estimate the transferability of neural networks of interest. Usable volumetric hydrogen capacity of MOFs under temperature-pressure swing conditions (77 K/100 bar and 160 K/5 bar) was implemented as an objective in the optimization task; 4146 structures from the CoRE MOF 2019 database and GCMC values from the dataset presented by Ahmed and Siegel\cite{ahmed2021predicting} were employed to train an ensemble of CG\textsuperscript{2}-SAGE models.

\subsection{Coarse-grained crystal graph processing}
The procedure outlined below was applied to construct the coarse-grained crystal graph and assign features to its nodes. Metals (in MOFs and COFs) and metalloids (in COFs) were completely removed from the initial crystal structure (the full list of atom types is provided in Figure S1). The removed atoms are regarded as metal centers, each of them was featurized using 200-dimensional matscholar embeddings\cite{weston2019named}. The crystal structure containing only nonmetal atoms was processed within OpenBabel routines\cite{o2011open} to produce Simplified Molecular Input Line Entry System\cite{weininger1988smiles,weininger1989smiles} (SMILES) strings from connected components in the reduced crystal graph. SMILES strings were then converted into 300-dimensional mol2vec embeddings\cite{jaeger2018mol2vec}. Metallic and organic units were built into the complete bipartite graph, where each node of one type is connected to a node of another type. In this form, the coarse-grained crystal graph was passed through a graph neural network.

\subsection{Model training}
CG\textsuperscript{2}-NNs were built with PyTorch\cite{paszke2019pytorch} and Deep Graph Library\cite{wang2019deep} (DGL). The choice of a deep learning framework has a substantial effect on energy efficiency of model training\cite{georgiou2022green}; therefore, PyTorch-based implementations were utilized for all other models as well. In the predictive performance (transfer learning) analysis, we used the Adam optimizer\cite{kingma2014adam} at a learning rate of ${10}^{-3}$ (${10}^{-4}$) and a batch size of 64 (16); the maximum number of epochs and the early stopping criterion were set to 500 and 50, respectively. Five-fold cross-validation was performed for model evaluation; one-eighth of training data was used for early stopping. To evaluate an objective in the optimization task (hydrogen storage capacity), an ensemble of 20 CG\textsuperscript{2}-SAGE models was trained by means of the Adam optimizer at a learning rate of ${10}^{-3}$ and a batch size of 64. The holdout cross-validation technique with a training/validation/test ratio of 80:10:10 was applied. The mean and standard deviation of model ensemble outputs were used to estimate the target property and the corresponding uncertainty\cite{lakshminarayanan2017simple}. The Eco2AI library\cite{budennyy2022eco2ai} was utilized to estimate GHG emissions of model training; the emission intensity coefficient was set to 240.56 kgCO\textsubscript{2}e/MW$\cdot$h (Moscow). All experiments in the study were conducted on a workstation equipped with two Intel® Xeon® CPUs E5-2695 v4 @ 2.10GHz, 144 GB RAM, and NVIDIA GeForce RTX 3090 Ti.

\subsection{Inverse reticular design}
The linker optimization algorithm was heavily inspired by previously developed methods\cite{bao2015silico1,bao2015silico2,nigam2021beyond}. Three symbolic operations (addition, deletion, and replacement) were applied to modify SELFIES strings\cite{krenn2020self}; each optimization step included one to 20 randomly selected mutations and several filtering procedures described below. The initial population was produced by generating 100,000 mutants from the SELFIES string corresponding to the parent structure (thieno[3,2-b]thiophene-2,5-dicarboxylic acid); 100 strings with the highest hydrogen storage capacity were subjected to further processing. Then, 1000 mutation steps were carried out. At each iteration, the population was updated by including a newly generated molecule if its objective was higher than the lowest value in the current generation; the number of molecules in the population was maintained at 100. Three filters were implemented to exclude irrelevant molecules. First, we assumed exactly two binding sites (carboxylic groups) in a molecule by applying the “linker-likeness” filter. Second, the molecules with low synthetic accessibility (SAscore\cite{ertl2009estimation} higher than 5.0) were also ignored. Third, 20 CG\textsuperscript{2}-SAGE models were used to assess uncertainty in predictions of the target property by means of deep-ensemble learning\cite{lakshminarayanan2017simple}. If the standard deviation of model ensemble outputs exceeded 2.5 g/l, then the corresponding molecule was discarded as well. One hundred launches of the optimization procedure were carried out in total.

The SELFIES strings included in the last generations were transferred into 3D atomic coordinates. In particular, the Experimental-Torsion basic Knowledge Distance Geometry\cite{riniker2015better} (ETKDG) approach implemented in the RDKit library was employed to generate 30 conformers for each of the 30 molecules with the highest hydrogen storage capacity. Then, local geometry optimization was performed using the ANI-2x neural-network potential\cite{devereux2020extending} and Broyden–Fletcher–Goldfarb–Shanno (BFGS) algorithm with the convergence criterion of ${10}^{-3}$ eV/Å for forces. Two quantities were calculated for each ensemble of conformers: 1) the mean angle between two vectors oriented from the carbon atom to the midpoint between oxygen atoms in carboxyl groups and 2) standard deviation of pairwise distance between carbon atoms in carboxyl groups. A molecule was included into the final set if the former value was higher than 150° and the latter one was less than 0.5 Å; the linker optimization was completed on four structures.

\section{Acknowledgements}
\label{sec:acknowledgements}
V.K. was supported by a Fellowship from Non-commercial Foundation for the Advancement of Science and Education INTELLECT.

\section{Code availability}
The PyTorch implementation of CG\textsuperscript{2}-NNs is available via PyPI (\url{https://pypi.org/project/cgcgnet}). The source code and examples are available at \url{https://github.com/korolewadim/cgcgnet}.

\section{Data availability}
All datasets used in the study are publicly available: band gap values and the corresponding MOF structures (\url{https://doi.org/10.6084/m9.figshare.13147324.v14}), thermal decomposition data (\url{https://doi.org/10.5281/zenodo.5737968}), CoRE MOF structures (\url{https://doi.org/10.5281/zenodo.7691378}), heat capacity data (\url{https://doi.org/10.24435/materialscloud:p1-2y}), adsorption data (\url{https://doi.org/10.24435/materialscloud:qt-cj}, \url{https://datahub.hymarc.org/dataset/computational-prediction-of-hydrogen-storage-capacities-in-mofs}), band gap values (\url{https://doi.org/10.5281/zenodo.7590815}) and the corresponding COF structures (\url{https://doi.org/10.24435/materialscloud:z6-jn}).

\bibliographystyle{unsrt}
\bibliography{references}

\begin{thebibliography}{100}

\bibitem{butler2018machine}
Keith~T Butler, Daniel~W Davies, Hugh Cartwright, Olexandr Isayev, and Aron Walsh.
\newblock Machine learning for molecular and materials science.
\newblock {\em Nature}, 559(7715):547--555, 2018.

\bibitem{schmidt2019recent}
Jonathan Schmidt, M{\'a}rio~RG Marques, Silvana Botti, and Miguel~AL Marques.
\newblock Recent advances and applications of machine learning in solid-state materials science.
\newblock {\em npj Computational Materials}, 5(1):83, 2019.

\bibitem{jablonka2020big}
Kevin~Maik Jablonka, Daniele Ongari, Seyed~Mohamad Moosavi, and Berend Smit.
\newblock Big-data science in porous materials: materials genomics and machine learning.
\newblock {\em Chemical reviews}, 120(16):8066--8129, 2020.

\bibitem{meredig2014combinatorial}
Bryce Meredig, Ankit Agrawal, Scott Kirklin, James~E Saal, Jeff~W Doak, Alan Thompson, Kunpeng Zhang, Alok Choudhary, and Christopher Wolverton.
\newblock Combinatorial screening for new materials in unconstrained composition space with machine learning.
\newblock {\em Physical Review B}, 89(9):094104, 2014.

\bibitem{schmidt2017predicting}
Jonathan Schmidt, Jingming Shi, Pedro Borlido, Liming Chen, Silvana Botti, and Miguel~AL Marques.
\newblock Predicting the thermodynamic stability of solids combining density functional theory and machine learning.
\newblock {\em Chemistry of Materials}, 29(12):5090--5103, 2017.

\bibitem{bartel2020critical}
Christopher~J Bartel, Amalie Trewartha, Qi~Wang, Alexander Dunn, Anubhav Jain, and Gerbrand Ceder.
\newblock A critical examination of compound stability predictions from machine-learned formation energies.
\newblock {\em npj computational materials}, 6(1):97, 2020.

\bibitem{lee2016prediction}
Joohwi Lee, Atsuto Seko, Kazuki Shitara, Keita Nakayama, and Isao Tanaka.
\newblock Prediction model of band gap for inorganic compounds by combination of density functional theory calculations and machine learning techniques.
\newblock {\em Physical Review B}, 93(11):115104, 2016.

\bibitem{zhuo2018predicting}
Ya~Zhuo, Aria Mansouri~Tehrani, and Jakoah Brgoch.
\newblock Predicting the band gaps of inorganic solids by machine learning.
\newblock {\em The journal of physical chemistry letters}, 9(7):1668--1673, 2018.

\bibitem{lu2018accelerated}
Shuaihua Lu, Qionghua Zhou, Yixin Ouyang, Yilv Guo, Qiang Li, and Jinlan Wang.
\newblock Accelerated discovery of stable lead-free hybrid organic-inorganic perovskites via machine learning.
\newblock {\em Nature communications}, 9(1):3405, 2018.

\bibitem{evans2017predicting}
Jack~D Evans and Fran{\c{c}}ois-Xavier Coudert.
\newblock Predicting the mechanical properties of zeolite frameworks by machine learning.
\newblock {\em Chemistry of Materials}, 29(18):7833--7839, 2017.

\bibitem{mansouri2018machine}
Aria Mansouri~Tehrani, Anton~O Oliynyk, Marcus Parry, Zeshan Rizvi, Samantha Couper, Feng Lin, Lowell Miyagi, Taylor~D Sparks, and Jakoah Brgoch.
\newblock Machine learning directed search for ultraincompressible, superhard materials.
\newblock {\em Journal of the American Chemical Society}, 140(31):9844--9853, 2018.

\bibitem{moghadam2019structure}
Peyman~Z Moghadam, Sven~MJ Rogge, Aurelia Li, Chun-Man Chow, Jelle Wieme, Noushin Moharrami, Marta Aragones-Anglada, Gareth Conduit, Diego~A Gomez-Gualdron, Veronique Van~Speybroeck, et~al.
\newblock Structure-mechanical stability relations of metal-organic frameworks via machine learning.
\newblock {\em Matter}, 1(1):219--234, 2019.

\bibitem{fernandez2014rapid}
Michael Fernandez, Peter~G Boyd, Thomas~D Daff, Mohammad~Zein Aghaji, and Tom~K Woo.
\newblock Rapid and accurate machine learning recognition of high performing metal organic frameworks for co2 capture.
\newblock {\em The journal of physical chemistry letters}, 5(17):3056--3060, 2014.

\bibitem{simon2015best}
Cory~M Simon, Rocio Mercado, Sondre~K Schnell, Berend Smit, and Maciej Haranczyk.
\newblock What are the best materials to separate a xenon/krypton mixture?
\newblock {\em Chemistry of Materials}, 27(12):4459--4475, 2015.

\bibitem{moosavi2020understanding}
Seyed~Mohamad Moosavi, Aditya Nandy, Kevin~Maik Jablonka, Daniele Ongari, Jon~Paul Janet, Peter~G Boyd, Yongjin Lee, Berend Smit, and Heather~J Kulik.
\newblock Understanding the diversity of the metal-organic framework ecosystem.
\newblock {\em Nature communications}, 11(1):1--10, 2020.

\bibitem{li2017high}
Zheng Li, Siwen Wang, Wei~Shan Chin, Luke~E Achenie, and Hongliang Xin.
\newblock High-throughput screening of bimetallic catalysts enabled by machine learning.
\newblock {\em Journal of Materials Chemistry A}, 5(46):24131--24138, 2017.

\bibitem{zahrt2019prediction}
Andrew~F Zahrt, Jeremy~J Henle, Brennan~T Rose, Yang Wang, William~T Darrow, and Scott~E Denmark.
\newblock Prediction of higher-selectivity catalysts by computer-driven workflow and machine learning.
\newblock {\em Science}, 363(6424):eaau5631, 2019.

\bibitem{chanussot2021open}
Lowik Chanussot, Abhishek Das, Siddharth Goyal, Thibaut Lavril, Muhammed Shuaibi, Morgane Riviere, Kevin Tran, Javier Heras-Domingo, Caleb Ho, Weihua Hu, et~al.
\newblock Open catalyst 2020 (oc20) dataset and community challenges.
\newblock {\em Acs Catalysis}, 11(10):6059--6072, 2021.

\bibitem{lecun2015deep}
Yann LeCun, Yoshua Bengio, and Geoffrey Hinton.
\newblock Deep learning.
\newblock {\em nature}, 521(7553):436--444, 2015.

\bibitem{xie2018crystal}
Tian Xie and Jeffrey~C Grossman.
\newblock Crystal graph convolutional neural networks for an accurate and interpretable prediction of material properties.
\newblock {\em Physical review letters}, 120(14):145301, 2018.

\bibitem{chen2019graph}
Chi Chen, Weike Ye, Yunxing Zuo, Chen Zheng, and Shyue~Ping Ong.
\newblock Graph networks as a universal machine learning framework for molecules and crystals.
\newblock {\em Chemistry of Materials}, 31(9):3564--3572, 2019.

\bibitem{korolev2019graph}
Vadim Korolev, Artem Mitrofanov, Alexandru Korotcov, and Valery Tkachenko.
\newblock Graph convolutional neural networks as “general-purpose” property predictors: the universality and limits of applicability.
\newblock {\em Journal of chemical information and modeling}, 60(1):22--28, 2020.

\bibitem{park2020developing}
Cheol~Woo Park and Chris Wolverton.
\newblock Developing an improved crystal graph convolutional neural network framework for accelerated materials discovery.
\newblock {\em Physical Review Materials}, 4(6):063801, 2020.

\bibitem{louis2020graph}
Steph-Yves Louis, Yong Zhao, Alireza Nasiri, Xiran Wang, Yuqi Song, Fei Liu, and Jianjun Hu.
\newblock Graph convolutional neural networks with global attention for improved materials property prediction.
\newblock {\em Physical Chemistry Chemical Physics}, 22(32):18141--18148, 2020.

\bibitem{karamad2020orbital}
Mohammadreza Karamad, Rishikesh Magar, Yuting Shi, Samira Siahrostami, Ian~D Gates, and Amir~Barati Farimani.
\newblock Orbital graph convolutional neural network for material property prediction.
\newblock {\em Physical Review Materials}, 4(9):093801, 2020.

\bibitem{cheng2021geometric}
Jiucheng Cheng, Chunkai Zhang, and Lifeng Dong.
\newblock A geometric-information-enhanced crystal graph network for predicting properties of materials.
\newblock {\em Communications Materials}, 2(1):92, 2021.

\bibitem{choudhary2021atomistic}
Kamal Choudhary and Brian DeCost.
\newblock Atomistic line graph neural network for improved materials property predictions.
\newblock {\em npj Computational Materials}, 7(1):185, 2021.

\bibitem{omee2022scalable}
Sadman~Sadeed Omee, Steph-Yves Louis, Nihang Fu, Lai Wei, Sourin Dey, Rongzhi Dong, Qinyang Li, and Jianjun Hu.
\newblock Scalable deeper graph neural networks for high-performance materials property prediction.
\newblock {\em Patterns}, 3(5):100491, 2022.

\bibitem{yan2022periodic}
Keqiang Yan, Yi~Liu, Yuchao Lin, and Shuiwang Ji.
\newblock Periodic graph transformers for crystal material property prediction.
\newblock {\em Advances in Neural Information Processing Systems}, 35:15066--15080, 2022.

\bibitem{jha2018elemnet}
Dipendra Jha, Logan Ward, Arindam Paul, Wei-keng Liao, Alok Choudhary, Chris Wolverton, and Ankit Agrawal.
\newblock Elemnet: Deep learning the chemistry of materials from only elemental composition.
\newblock {\em Scientific reports}, 8(1):17593, 2018.

\bibitem{goodall2020predicting}
Rhys~EA Goodall and Alpha~A Lee.
\newblock Predicting materials properties without crystal structure: Deep representation learning from stoichiometry.
\newblock {\em Nature communications}, 11(1):6280, 2020.

\bibitem{wang2021compositionally}
Anthony Yu-Tung Wang, Steven~K Kauwe, Ryan~J Murdock, and Taylor~D Sparks.
\newblock Compositionally restricted attention-based network for materials property predictions.
\newblock {\em Npj Computational Materials}, 7(1):77, 2021.

\bibitem{faber2016machine}
Felix~A Faber, Alexander Lindmaa, O~Anatole Von~Lilienfeld, and Rickard Armiento.
\newblock Machine learning energies of 2 million elpasolite (a b c 2 d 6) crystals.
\newblock {\em Physical review letters}, 117(13):135502, 2016.

\bibitem{ye2018deep}
Weike Ye, Chi Chen, Zhenbin Wang, Iek-Heng Chu, and Shyue~Ping Ong.
\newblock Deep neural networks for accurate predictions of crystal stability.
\newblock {\em Nature communications}, 9(1):3800, 2018.

\bibitem{balachandran2018predictions}
Prasanna~V Balachandran, Antoine~A Emery, James~E Gubernatis, Turab Lookman, Chris Wolverton, and Alex Zunger.
\newblock Predictions of new ab o 3 perovskite compounds by combining machine learning and density functional theory.
\newblock {\em Physical Review Materials}, 2(4):043802, 2018.

\bibitem{yaghi2003reticular}
Omar~M Yaghi, Michael O'Keeffe, Nathan~W Ockwig, Hee~K Chae, Mohamed Eddaoudi, and Jaheon Kim.
\newblock Reticular synthesis and the design of new materials.
\newblock {\em Nature}, 423(6941):705--714, 2003.

\bibitem{lyu2020digital}
Hao Lyu, Zhe Ji, Stefan Wuttke, and Omar~M Yaghi.
\newblock Digital reticular chemistry.
\newblock {\em Chem}, 6(9):2219--2241, 2020.

\bibitem{li1999design}
Hailian Li, Mohamed Eddaoudi, Michael O'Keeffe, and Omar~M Yaghi.
\newblock Design and synthesis of an exceptionally stable and highly porous metal-organic framework.
\newblock {\em nature}, 402(6759):276--279, 1999.

\bibitem{cote2005porous}
Adrien~P Cote, Annabelle~I Benin, Nathan~W Ockwig, Michael O'Keeffe, Adam~J Matzger, and Omar~M Yaghi.
\newblock Porous, crystalline, covalent organic frameworks.
\newblock {\em science}, 310(5751):1166--1170, 2005.

\bibitem{wang2018sensing}
Hao Wang, William~P Lustig, and Jing Li.
\newblock Sensing and capture of toxic and hazardous gases and vapors by metal--organic frameworks.
\newblock {\em Chemical Society Reviews}, 47(13):4729--4756, 2018.

\bibitem{lin2019exploration}
Rui-Biao Lin, Shengchang Xiang, Huabin Xing, Wei Zhou, and Banglin Chen.
\newblock Exploration of porous metal--organic frameworks for gas separation and purification.
\newblock {\em Coordination chemistry reviews}, 378:87--103, 2019.

\bibitem{zhu2017metal}
Li~Zhu, Xiao-Qin Liu, Hai-Long Jiang, and Lin-Bing Sun.
\newblock Metal--organic frameworks for heterogeneous basic catalysis.
\newblock {\em Chemical reviews}, 117(12):8129--8176, 2017.

\bibitem{huang2017multifunctional}
Yuan-Biao Huang, Jun Liang, Xu-Sheng Wang, and Rong Cao.
\newblock Multifunctional metal--organic framework catalysts: synergistic catalysis and tandem reactions.
\newblock {\em Chemical Society Reviews}, 46(1):126--157, 2017.

\bibitem{lee2021computational}
Sangwon Lee, Baekjun Kim, Hyun Cho, Hooseung Lee, Sarah~Yunmi Lee, Eun~Seon Cho, and Jihan Kim.
\newblock Computational screening of trillions of metal--organic frameworks for high-performance methane storage.
\newblock {\em ACS Applied Materials \& Interfaces}, 13(20):23647--23654, 2021.

\bibitem{yao2021inverse}
Zhenpeng Yao, Benjam{\'\i}n S{\'a}nchez-Lengeling, N~Scott Bobbitt, Benjamin~J Bucior, Sai Govind~Hari Kumar, Sean~P Collins, Thomas Burns, Tom~K Woo, Omar~K Farha, Randall~Q Snurr, et~al.
\newblock Inverse design of nanoporous crystalline reticular materials with deep generative models.
\newblock {\em Nature Machine Intelligence}, 3(1):76--86, 2021.

\bibitem{chen2022interpretable}
Pin Chen, Rui Jiao, Jinyu Liu, Yang Liu, and Yutong Lu.
\newblock Interpretable graph transformer network for predicting adsorption isotherms of metal--organic frameworks.
\newblock {\em Journal of Chemical Information and Modeling}, 62(22):5446--5456, 2022.

\bibitem{kang2023multi}
Yeonghun Kang, Hyunsoo Park, Berend Smit, and Jihan Kim.
\newblock A multi-modal pre-training transformer for universal transfer learning in metal--organic frameworks.
\newblock {\em Nature Machine Intelligence}, 5(3):309--318, 2023.

\bibitem{cao2023moformer}
Zhonglin Cao, Rishikesh Magar, Yuyang Wang, and Amir Barati~Farimani.
\newblock Moformer: self-supervised transformer model for metal--organic framework property prediction.
\newblock {\em Journal of the American Chemical Society}, 145(5):2958--2967, 2023.

\bibitem{kim2020inverse}
Baekjun Kim, Sangwon Lee, and Jihan Kim.
\newblock Inverse design of porous materials using artificial neural networks.
\newblock {\em Science advances}, 6(1):eaax9324, 2020.

\bibitem{park2022computational}
Junkil Park, Yunsung Lim, Sangwon Lee, and Jihan Kim.
\newblock Computational design of metal--organic frameworks with unprecedented high hydrogen working capacity and high synthesizability.
\newblock {\em Chemistry of Materials}, 35(1):9--16, 2023.

\bibitem{reiser2022graph}
Patrick Reiser, Marlen Neubert, Andr{\'e} Eberhard, Luca Torresi, Chen Zhou, Chen Shao, Houssam Metni, Clint van Hoesel, Henrik Schopmans, Timo Sommer, et~al.
\newblock Graph neural networks for materials science and chemistry.
\newblock {\em Communications Materials}, 3(1):93, 2022.

\bibitem{gilmer2017neural}
Justin Gilmer, Samuel~S Schoenholz, Patrick~F Riley, Oriol Vinyals, and George~E Dahl.
\newblock Neural message passing for quantum chemistry.
\newblock In {\em International conference on machine learning}, pages 1263--1272. PMLR, 2017.

\bibitem{alsentzer2020subgraph}
Emily Alsentzer, Samuel Finlayson, Michelle Li, and Marinka Zitnik.
\newblock Subgraph neural networks.
\newblock {\em Advances in Neural Information Processing Systems}, 33:8017--8029, 2020.

\bibitem{sun2021sugar}
Qingyun Sun, Jianxin Li, Hao Peng, Jia Wu, Yuanxing Ning, Philip~S Yu, and Lifang He.
\newblock Sugar: Subgraph neural network with reinforcement pooling and self-supervised mutual information mechanism.
\newblock In {\em Proceedings of the Web Conference 2021}, pages 2081--2091, 2021.

\bibitem{barthel2018distinguishing}
Senja Barthel, Eugeny~V Alexandrov, Davide~M Proserpio, and Berend Smit.
\newblock Distinguishing metal--organic frameworks.
\newblock {\em Crystal growth \& design}, 18(3):1738--1747, 2018.

\bibitem{chung2014computation}
Yongchul~G Chung, Jeffrey Camp, Maciej Haranczyk, Benjamin~J Sikora, Wojciech Bury, Vaiva Krungleviciute, Taner Yildirim, Omar~K Farha, David~S Sholl, and Randall~Q Snurr.
\newblock Computation-ready, experimental metal--organic frameworks: A tool to enable high-throughput screening of nanoporous crystals.
\newblock {\em Chemistry of Materials}, 26(21):6185--6192, 2014.

\bibitem{chung2019advances}
Yongchul~G Chung, Emmanuel Haldoupis, Benjamin~J Bucior, Maciej Haranczyk, Seulchan Lee, Hongda Zhang, Konstantinos~D Vogiatzis, Marija Milisavljevic, Sanliang Ling, Jeffrey~S Camp, et~al.
\newblock Advances, updates, and analytics for the computation-ready, experimental metal--organic framework database: Core mof 2019.
\newblock {\em Journal of Chemical \& Engineering Data}, 64(12):5985--5998, 2019.

\bibitem{bucior2019identification}
Benjamin~J Bucior, Andrew~S Rosen, Maciej Haranczyk, Zhenpeng Yao, Michael~E Ziebel, Omar~K Farha, Joseph~T Hupp, J~Ilja Siepmann, Al{\'a}n Aspuru-Guzik, and Randall~Q Snurr.
\newblock Identification schemes for metal--organic frameworks to enable rapid search and cheminformatics analysis.
\newblock {\em Crystal Growth \& Design}, 19(11):6682--6697, 2019.

\bibitem{jablonka2023ecosystem}
Kevin~Maik Jablonka, Andrew~S Rosen, Aditi~S Krishnapriyan, and Berend Smit.
\newblock An ecosystem for digital reticular chemistry.
\newblock {\em ACS Central Science}, 9(4):563--581, 2023.

\bibitem{jaeger2018mol2vec}
Sabrina Jaeger, Simone Fulle, and Samo Turk.
\newblock Mol2vec: unsupervised machine learning approach with chemical intuition.
\newblock {\em Journal of chemical information and modeling}, 58(1):27--35, 2018.

\bibitem{weston2019named}
Leigh Weston, Vahe Tshitoyan, John Dagdelen, Olga Kononova, Amalie Trewartha, Kristin~A Persson, Gerbrand Ceder, and Anubhav Jain.
\newblock Named entity recognition and normalization applied to large-scale information extraction from the materials science literature.
\newblock {\em Journal of chemical information and modeling}, 59(9):3692--3702, 2019.

\bibitem{nicholas2020understanding}
Thomas~C Nicholas, Andrew~L Goodwin, and Volker~L Deringer.
\newblock Understanding the geometric diversity of inorganic and hybrid frameworks through structural coarse-graining.
\newblock {\em Chemical Science}, 11(46):12580--12587, 2020.

\bibitem{nicholas2021visualization}
Thomas~C Nicholas, Eugeny~V Alexandrov, Vladislav~A Blatov, Alexander~P Shevchenko, Davide~M Proserpio, Andrew~L Goodwin, and Volker~L Deringer.
\newblock Visualization and quantification of geometric diversity in metal--organic frameworks.
\newblock {\em Chemistry of Materials}, 33(21):8289--8300, 2021.

\bibitem{beaulieu2023coarse}
Zo{\'e}~Faure Beaulieu, Thomas~C Nicholas, John~LA Gardner, Andrew~L Goodwin, and Volker~L Deringer.
\newblock Coarse-grained versus fully atomistic machine learning for zeolitic imidazolate frameworks.
\newblock {\em Chemical Communications}, 59(76):11405--11408, 2023.

\bibitem{ba2016layer}
Jimmy~Lei Ba, Jamie~Ryan Kiros, and Geoffrey~E Hinton.
\newblock Layer normalization.
\newblock {\em arXiv preprint arXiv:1607.06450}, 2016.

\bibitem{clevert2015fast}
Djork-Arn{\'e} Clevert, Thomas Unterthiner, and Sepp Hochreiter.
\newblock Fast and accurate deep network learning by exponential linear units (elus).
\newblock {\em arXiv preprint arXiv:1511.07289}, 2015.

\bibitem{kipf2016semi}
Thomas~N Kipf and Max Welling.
\newblock Semi-supervised classification with graph convolutional networks.
\newblock {\em arXiv preprint arXiv:1609.02907}, 2016.

\bibitem{velivckovic2017graph}
Petar Veli{\v{c}}kovi{\'c}, Guillem Cucurull, Arantxa Casanova, Adriana Romero, Pietro Lio, and Yoshua Bengio.
\newblock Graph attention networks.
\newblock {\em arXiv preprint arXiv:1710.10903}, 2017.

\bibitem{hamilton2017inductive}
Will Hamilton, Zhitao Ying, and Jure Leskovec.
\newblock Inductive representation learning on large graphs.
\newblock {\em Advances in neural information processing systems}, 30, 2017.

\bibitem{brody2021attentive}
Shaked Brody, Uri Alon, and Eran Yahav.
\newblock How attentive are graph attention networks?
\newblock {\em arXiv preprint arXiv:2105.14491}, 2021.

\bibitem{rosen2021machine}
Andrew~S Rosen, Shaelyn~M Iyer, Debmalya Ray, Zhenpeng Yao, Alan Aspuru-Guzik, Laura Gagliardi, Justin~M Notestein, and Randall~Q Snurr.
\newblock Machine learning the quantum-chemical properties of metal--organic frameworks for accelerated materials discovery.
\newblock {\em Matter}, 4(5):1578--1597, 2021.

\bibitem{rosen2022high}
Andrew~S Rosen, Victor Fung, Patrick Huck, Cody~T O’Donnell, Matthew~K Horton, Donald~G Truhlar, Kristin~A Persson, Justin~M Notestein, and Randall~Q Snurr.
\newblock High-throughput predictions of metal--organic framework electronic properties: theoretical challenges, graph neural networks, and data exploration.
\newblock {\em npj Computational Materials}, 8(1):112, 2022.

\bibitem{li2021launch}
Aurelia Li, Rocio~Bueno Perez, Seth Wiggin, Suzanna~C Ward, Peter~A Wood, and David Fairen-Jimenez.
\newblock The launch of a freely accessible mof cif collection from the csd.
\newblock {\em Matter}, 4(4):1105--1106, 2021.

\bibitem{goodall2022rapid}
Rhys~EA Goodall, Abhijith~S Parackal, Felix~A Faber, Rickard Armiento, and Alpha~A Lee.
\newblock Rapid discovery of stable materials by coordinate-free coarse graining.
\newblock {\em Science Advances}, 8(30):eabn4117, 2022.

\bibitem{alexander2015beware}
David~LJ Alexander, Alexander Tropsha, and David~A Winkler.
\newblock Beware of r2: simple, unambiguous assessment of the prediction accuracy of qsar and qspr models.
\newblock {\em Journal of chemical information and modeling}, 55(7):1316--1322, 2015.

\bibitem{healy2020thermal}
Colm Healy, Komal~M Patil, Benjamin~H Wilson, Lily Hermanspahn, Nathan~C Harvey-Reid, Ben~I Howard, Carline Kleinjan, James Kolien, Fabian Payet, Shane~G Telfer, et~al.
\newblock The thermal stability of metal-organic frameworks.
\newblock {\em Coordination Chemistry Reviews}, 419:213388, 2020.

\bibitem{yamada2019predicting}
Hironao Yamada, Chang Liu, Stephen Wu, Yukinori Koyama, Shenghong Ju, Junichiro Shiomi, Junko Morikawa, and Ryo Yoshida.
\newblock Predicting materials properties with little data using shotgun transfer learning.
\newblock {\em ACS central science}, 5(10):1717--1730, 2019.

\bibitem{jha2019enhancing}
Dipendra Jha, Kamal Choudhary, Francesca Tavazza, Wei-keng Liao, Alok Choudhary, Carelyn Campbell, and Ankit Agrawal.
\newblock Enhancing materials property prediction by leveraging computational and experimental data using deep transfer learning.
\newblock {\em Nature communications}, 10(1):5316, 2019.

\bibitem{gupta2021cross}
Vishu Gupta, Kamal Choudhary, Francesca Tavazza, Carelyn Campbell, Wei-keng Liao, Alok Choudhary, and Ankit Agrawal.
\newblock Cross-property deep transfer learning framework for enhanced predictive analytics on small materials data.
\newblock {\em Nature communications}, 12(1):6595, 2021.

\bibitem{suzuki2022self}
Yuta Suzuki, Tatsunori Taniai, Kotaro Saito, Yoshitaka Ushiku, and Kanta Ono.
\newblock Self-supervised learning of materials concepts from crystal structures via deep neural networks.
\newblock {\em Machine Learning: Science and Technology}, 3(4):045034, 2022.

\bibitem{korolev2023accurate}
Vadim Korolev and Pavel Protsenko.
\newblock Accurate, interpretable predictions of materials properties within transformer language models.
\newblock {\em Patterns}, 4(10):100803, 2023.

\bibitem{mcinnes2018umap}
Leland McInnes, John Healy, and James Melville.
\newblock Umap: Uniform manifold approximation and projection for dimension reduction.
\newblock {\em arXiv preprint arXiv:1802.03426}, 2018.

\bibitem{probst2023aiming}
Daniel Probst.
\newblock Aiming beyond slight increases in accuracy.
\newblock {\em Nature Reviews Chemistry}, 7(4):227--228, 2023.

\bibitem{korolev2023carbon}
Vadim Korolev and Artem Mitrofanov.
\newblock Carbon footprint of artificial intelligence in materials science: Should we be concerned?
\newblock {\em ChemRxiv}, 2023.

\bibitem{rowsell2006effects}
Jesse~LC Rowsell and Omar~M Yaghi.
\newblock Effects of functionalization, catenation, and variation of the metal oxide and organic linking units on the low-pressure hydrogen adsorption properties of metal- organic frameworks.
\newblock {\em Journal of the American Chemical Society}, 128(4):1304--1315, 2006.

\bibitem{krenn2020self}
Mario Krenn, Florian H{\"a}se, AkshatKumar Nigam, Pascal Friederich, and Alan Aspuru-Guzik.
\newblock Self-referencing embedded strings (selfies): A 100\% robust molecular string representation.
\newblock {\em Machine Learning: Science and Technology}, 1(4):045024, 2020.

\bibitem{lakshminarayanan2017simple}
Balaji Lakshminarayanan, Alexander Pritzel, and Charles Blundell.
\newblock Simple and scalable predictive uncertainty estimation using deep ensembles.
\newblock {\em Advances in neural information processing systems}, 30, 2017.

\bibitem{korolev2022universal}
Vadim Korolev, Iurii Nevolin, and Pavel Protsenko.
\newblock A universal similarity based approach for predictive uncertainty quantification in materials science.
\newblock {\em Scientific Reports}, 12(1):14931, 2022.

\bibitem{ahmed2017balancing}
Alauddin Ahmed, Yiyang Liu, Justin Purewal, Ly~D Tran, Antek~G Wong-Foy, Mike Veenstra, Adam~J Matzger, and Donald~J Siegel.
\newblock Balancing gravimetric and volumetric hydrogen density in mofs.
\newblock {\em Energy \& Environmental Science}, 10(11):2459--2471, 2017.

\bibitem{luo2022mof}
Yi~Luo, Saientan Bag, Orysia Zaremba, Adrian Cierpka, Jacopo Andreo, Stefan Wuttke, Pascal Friederich, and Manuel Tsotsalas.
\newblock Mof synthesis prediction enabled by automatic data mining and machine learning.
\newblock {\em Angewandte Chemie International Edition}, 61(19):e202200242, 2022.

\bibitem{zheng2023chatgpt}
Zhiling Zheng, Oufan Zhang, Christian Borgs, Jennifer~T Chayes, and Omar~M Yaghi.
\newblock Chatgpt chemistry assistant for text mining and the prediction of mof synthesis.
\newblock {\em Journal of the American Chemical Society}, 145(32):18048--18062, 2023.

\bibitem{aulakh2015importance}
Darpandeep Aulakh, Juby~R Varghese, and Mario Wriedt.
\newblock The importance of polymorphism in metal--organic framework studies.
\newblock {\em Inorganic Chemistry}, 54(17):8679--8684, 2015.

\bibitem{widmer2019rich}
Remo~N Widmer, Giulio~I Lampronti, Siwar Chibani, Craig~W Wilson, Simone Anzellini, Stefan Farsang, Annette~K Kleppe, Nicola~PM Casati, Simon~G MacLeod, Simon~AT Redfern, et~al.
\newblock Rich polymorphism of a metal--organic framework in pressure--temperature space.
\newblock {\em Journal of the American Chemical Society}, 141(23):9330--9337, 2019.

\bibitem{zhou2022inverse}
Musen Zhou and Jianzhong Wu.
\newblock Inverse design of metal--organic frameworks for c2h4/c2h6 separation.
\newblock {\em npj Computational Materials}, 8(1):256, 2022.

\bibitem{comlek2023rapid}
Yigitcan Comlek, Thang~Duc Pham, Randall~Q Snurr, and Wei Chen.
\newblock Rapid design of top-performing metal-organic frameworks with qualitative representations of building blocks.
\newblock {\em npj Computational Materials}, 9(1):170, 2023.

\bibitem{lim2021finely}
Yunsung Lim, Junkil Park, Sangwon Lee, and Jihan Kim.
\newblock Finely tuned inverse design of metal--organic frameworks with user-desired xe/kr selectivity.
\newblock {\em Journal of Materials Chemistry A}, 9(37):21175--21183, 2021.

\bibitem{perdew1996generalized}
John~P Perdew, Kieron Burke, and Matthias Ernzerhof.
\newblock Generalized gradient approximation made simple.
\newblock {\em Physical review letters}, 77(18):3865, 1996.

\bibitem{verma2017hle17}
Pragya Verma and Donald~G Truhlar.
\newblock Hle17: An improved local exchange--correlation functional for computing semiconductor band gaps and molecular excitation energies.
\newblock {\em The Journal of Physical Chemistry C}, 121(13):7144--7154, 2017.

\bibitem{krukau2006influence}
Aliaksandr~V Krukau, Oleg~A Vydrov, Artur~F Izmaylov, and Gustavo~E Scuseria.
\newblock Influence of the exchange screening parameter on the performance of screened hybrid functionals.
\newblock {\em The Journal of chemical physics}, 125(22), 2006.

\bibitem{nandy2022mofsimplify}
Aditya Nandy, Gianmarco Terrones, Naveen Arunachalam, Chenru Duan, David~W Kastner, and Heather~J Kulik.
\newblock Mofsimplify, machine learning models with extracted stability data of three thousand metal--organic frameworks.
\newblock {\em Scientific Data}, 9(1):74, 2022.

\bibitem{moosavi2022data}
Seyed~Mohamad Moosavi, Bal{\'a}zs~{\'A}lmos Novotny, Daniele Ongari, Elias Moubarak, Mehrdad Asgari, {\"O}zge Kadioglu, Charithea Charalambous, Andres Ortega-Guerrero, Amir~H Farmahini, Lev Sarkisov, et~al.
\newblock A data-science approach to predict the heat capacity of nanoporous materials.
\newblock {\em Nature materials}, 21(12):1419--1425, 2022.

\bibitem{mourino2023search}
Beatriz Mourino, Kevin~Maik Jablonka, Andres Ortega-Guerrero, and Berend Smit.
\newblock In search of covalent organic framework photocatalysts: A dft-based screening approach.
\newblock {\em Advanced Functional Materials}, 33(32):2301594, 2023.

\bibitem{ongari2019building}
Daniele Ongari, Aliaksandr~V Yakutovich, Leopold Talirz, and Berend Smit.
\newblock Building a consistent and reproducible database for adsorption evaluation in covalent--organic frameworks.
\newblock {\em ACS central science}, 5(10):1663--1675, 2019.

\bibitem{ahmed2021predicting}
Alauddin Ahmed and Donald~J Siegel.
\newblock Predicting hydrogen storage in mofs via machine learning.
\newblock {\em Patterns}, 2(7), 2021.

\bibitem{o2011open}
Noel~M O'Boyle, Michael Banck, Craig~A James, Chris Morley, Tim Vandermeersch, and Geoffrey~R Hutchison.
\newblock Open babel: An open chemical toolbox.
\newblock {\em Journal of cheminformatics}, 3(1):1--14, 2011.

\bibitem{weininger1988smiles}
David Weininger.
\newblock Smiles, a chemical language and information system. 1. introduction to methodology and encoding rules.
\newblock {\em Journal of chemical information and computer sciences}, 28(1):31--36, 1988.

\bibitem{weininger1989smiles}
David Weininger, Arthur Weininger, and Joseph~L Weininger.
\newblock Smiles. 2. algorithm for generation of unique smiles notation.
\newblock {\em Journal of chemical information and computer sciences}, 29(2):97--101, 1989.

\bibitem{paszke2019pytorch}
Adam Paszke, Sam Gross, Francisco Massa, Adam Lerer, James Bradbury, Gregory Chanan, Trevor Killeen, Zeming Lin, Natalia Gimelshein, Luca Antiga, et~al.
\newblock Pytorch: An imperative style, high-performance deep learning library.
\newblock {\em Advances in neural information processing systems}, 32, 2019.

\bibitem{wang2019deep}
Minjie Wang, Da~Zheng, Zihao Ye, Quan Gan, Mufei Li, Xiang Song, Jinjing Zhou, Chao Ma, Lingfan Yu, Yu~Gai, et~al.
\newblock Deep graph library: A graph-centric, highly-performant package for graph neural networks.
\newblock {\em arXiv preprint arXiv:1909.01315}, 2019.

\bibitem{georgiou2022green}
Stefanos Georgiou, Maria Kechagia, Tushar Sharma, Federica Sarro, and Ying Zou.
\newblock Green ai: Do deep learning frameworks have different costs?
\newblock In {\em Proceedings of the 44th International Conference on Software Engineering}, pages 1082--1094, 2022.

\bibitem{kingma2014adam}
Diederik~P Kingma and Jimmy Ba.
\newblock Adam: A method for stochastic optimization.
\newblock {\em arXiv preprint arXiv:1412.6980}, 2014.

\bibitem{budennyy2022eco2ai}
Semen~Andreevich Budennyy, Vladimir~Dmitrievich Lazarev, Nikita~Nikolaevich Zakharenko, Aleksei~N Korovin, OA~Plosskaya, Denis~Valer'evich Dimitrov, VS~Akhripkin, IV~Pavlov, Ivan~Valer'evich Oseledets, Ivan~Segundovich Barsola, et~al.
\newblock Eco2ai: carbon emissions tracking of machine learning models as the first step towards sustainable ai.
\newblock In {\em Doklady Mathematics}, volume 106, pages S118--S128. Springer, 2022.

\bibitem{bao2015silico1}
Yi~Bao, Richard~L Martin, Cory~M Simon, Maciej Haranczyk, Berend Smit, and Michael~W Deem.
\newblock In silico discovery of high deliverable capacity metal--organic frameworks.
\newblock {\em The Journal of Physical Chemistry C}, 119(1):186--195, 2015.

\bibitem{bao2015silico2}
Yi~Bao, Richard~L Martin, Maciej Haranczyk, and Michael~W Deem.
\newblock In silico prediction of mofs with high deliverable capacity or internal surface area.
\newblock {\em Physical Chemistry Chemical Physics}, 17(18):11962--11973, 2015.

\bibitem{nigam2021beyond}
AkshatKumar Nigam, Robert Pollice, Mario Krenn, Gabriel dos Passos~Gomes, and Alan Aspuru-Guzik.
\newblock Beyond generative models: superfast traversal, optimization, novelty, exploration and discovery (stoned) algorithm for molecules using selfies.
\newblock {\em Chemical science}, 12(20):7079--7090, 2021.

\bibitem{ertl2009estimation}
Peter Ertl and Ansgar Schuffenhauer.
\newblock Estimation of synthetic accessibility score of drug-like molecules based on molecular complexity and fragment contributions.
\newblock {\em Journal of cheminformatics}, 1:1--11, 2009.

\bibitem{riniker2015better}
Sereina Riniker and Gregory~A Landrum.
\newblock Better informed distance geometry: using what we know to improve conformation generation.
\newblock {\em Journal of chemical information and modeling}, 55(12):2562--2574, 2015.

\bibitem{devereux2020extending}
Christian Devereux, Justin~S Smith, Kate~K Huddleston, Kipton Barros, Roman Zubatyuk, Olexandr Isayev, and Adrian~E Roitberg.
\newblock Extending the applicability of the ani deep learning molecular potential to sulfur and halogens.
\newblock {\em Journal of Chemical Theory and Computation}, 16(7):4192--4202, 2020.

\end{thebibliography}

\end{document}